\documentclass[preprintnumbers,amsmath,11pt,amssymb,floatfix,superscriptaddress,showpacs,nofootinbib]{revtex4}
\usepackage{graphicx}
\usepackage{epsfig}
\usepackage{bm}
\usepackage{amsfonts}

\begin{document}

\title{\Large QCD ghost reconstruction of $f(T)$ gravity in flat FRW universe}

\author{Surajit Chattopadhyay}
\affiliation{ Department of Computer Application, Pailan College
of Management and Technology, Bengal Pailan Park, Kolkata-700 104,
India.} \footnote{$^{,}$$^{*}$ Corresponding author, email:
surajitchatto@outlook.com, surajcha@iucaa.ernet.in}

\begin{abstract}
\textbf{Abstract:}The present work reports a reconstruction scheme for $f(T)$ gravity based on QCD ghost dark energy. Two models of $f(T)$ have been generated and the pressure and density contributions due to torsion have been reconstructed. Two realistic models have been obtained and the effective equations of state have been studied. Also, the squared speed of sound has been studied to examine the stability of the models.
\end{abstract}

\pacs{98.80.-k; 04.50.Kd}

\maketitle

\section{Introduction}
Accelerated expansion of our universe, as evidenced by Supernovae Ia (SNeIa),
Cosmic Microwave Background (CMB) radiation anisotropies,
Large Scale Structure (LSS) and X-ray experiments, is well documented in literature \cite{obs1,obs2}. A missing energy component, also known
as Dark Energy (DE) (reviewed in \cite{DE1,DE2,DE3,DE4,DE5,DE6,DE7}) with negative pressure, is widely considered by
scientists as responsible of this accelerated expansion. The simplest model of DE is the cosmological constant, which is a key
ingredient in the $\Lambda$CDM model. Although the $\Lambda$CDM model is consistent very well with
all observational data, it has the fine tuning problem. Plenty of other DE models
have been proposed till date (see \cite{DE6} and references therein), but almost all of them explain the acceleration
expansion either by introducing new degree(s) of freedom or by modifying gravity \cite{gde1}. Evolution of DE parameter within scope of a spatially homogeneous and isotropic Friedmann-Robertson-Walker (FRW) model filled with perfect fluid and dark energy components has been studied in \cite{DE8} by generalizing recent results. Some well-known cosmological parameters in the framework of interacting generalized holographic dark energy with cold dark matter in non-flat FRW universe have been determined in the work of \cite{DE9}.

A DE model, so-called Veneziano ghost DE (GDE), has been proposed in \cite{gde2}. The key ingredient of this new model is that the Veneziano ghost, which is unphysical in the usual Minkowski spacetime quantum field theory (QFT), exhibits important physical effects in dynamical
spacetime or spacetime with non-trivial topology. Veneziano ghost is supposed to exist for
solving the $U(1)$ problem in low-energy effective theory of QCD \cite{gde1}. Although in
flat Minkowski spacetime the QCD ghosts are unphysical and make no contribution, in curved/time-dependent
backgrounds the cancellation of their contribution to the vacuum energy leave a small energy density \cite{QCD1} $\rho\sim \Lambda_{QCD}^3$, where $H$ is the Hubble parameter and $\Lambda_{QCD}\sim 100 MeV$ is the QCD mass scale.

In the present work, our purpose is to reconstruct $f(T)$ gravity based on QCD GDE. The $f(T)$ ($T$ is torsion) gravity
is an interesting sort of modified theories of gravity reviewed in \cite{Nojiri1,Nojiri3}. Various aspects of $f(T)$ gravity have been discussed in references \cite{momeni1} to \cite{myrza1}. Reconstruction of modified gravity and dark energy is not new. Reconstruction schemes for dark energy models have been attempted in \cite{setare1,setare2,setare3,setare4,setare5,setare6,ijp1}. The studies that are more relevant to the present work fall in the category of DE based reconstruction of modified gravity model. One remarkable $f(T)$ gravity reconstruction work is \cite{recons1}, which demonstrated that there appear finite-time future singularities in $f(T)$ gravity with $T$ being the torsion scalar. Ref. \cite{recons1} reconstructed a model of $f(T)$ gravity with realizing the finite-time future singularities. In two difference works, refs. \cite{recons2} and \cite{recons3} demonstrated holographic reconstruction of $f(T)$ gravity. Ref. \cite{recons2} showed that the evolutionary nature of the holographic dark energy is essentially based on two important parameters, $\Omega_V$  and $w_V$ ,respectively, the dimensionless dark energy and the parameter of the equation of state, related to the holographic dark energy. On the other hand, ref. \cite{recons3} derived two alternative holographic solutions for $f(T)$ and studied their stability through the squared speed of sound $v_s^2$. The current work is largely motivated by the work of the reference \cite{recons4}, that investigated cosmological application of holographic dark energy density in the modified gravity framework and employed the holographic model of dark energy to obtain the equation of state for the holographic energy density in a spatially flat universe.

In the current work we shall reconstruct $f(T)$ gravity based on QCD ghost dark energy and investigate its cosmological consequences. In Section II we shall discuss the reconstruction scheme along with discussions on the plots. In Section III we shall present the crisp outcomes.

\section{The reconstruction scheme}
In this section we shall present a reconstruction scheme for $f(T)$ gravity. For that purpose, our choice for scale factor in the present work is
\begin{equation}\label{a}
a=a_0 t^{n}
\end{equation}
Hence, the Hubble parameter gets the form $H=\frac{n}{t}$. We consider the GDE whose energy density is proportional to the Hubble parameter \cite{QCD1}
\begin{equation}\label{qde}
\rho_{gde}=\frac{\alpha  (1-\epsilon )}{\tilde{r}_h}=\alpha  (1-\epsilon )\sqrt{H^2+\frac{k}{a^2}};~~\epsilon\equiv\frac{\dot{\tilde{r}}_h}{2H\tilde{r}_h}
\end{equation}
Here, $\alpha$ is a constant with dimension $(energy)^3$ and roughly of order of $\Lambda_{QCD}^3$, where $\Lambda_{QCD}\sim 100 MeV$.  If we ignore the spatial curvature, as we do in this paper, the
trapping horizon is coincident with the Hubble horizon $\tilde{r}_h=1/H$, and
\begin{equation}\label{qdeH}
\rho_{gde}=\alpha  (1-\epsilon ) H;~~\epsilon=-\dot{H}/2H
\end{equation}
We shall reconstruct $f(T)$ gravity for the QCD GDE given in Eq. (\ref{qdeH}).
In the framework of $f(T)$ theory, the action of modified
teleparallel action is given by
\begin{equation}
I=\frac{1}{16\pi G}\int d^{4}x \sqrt{-g}\left[f(T)+L_{m}\right],
\label{1}
\end{equation}
where $L_{m}$ is the Lagrangian density of the matter inside the
universe, $G$ is the gravitational constant and $g$ is the determinant of the metric tensor $g^{\mu \nu}$. We consider a flat Friedmann-Robertson-Walker (FRW)
universe filled with the pressureless matter. Choosing $(8\pi
G=1)$, modified Friedmann equations in the framework of $f(T)$
gravity are given by
\begin{eqnarray}
H^{2}&=&\frac{1}{3}\left(\rho+\rho_{T}\right), \label{2}\\
2\dot{H}+3H^{2}&=&-\left(p+p_{T}\right), \label{3}
\end{eqnarray}
where
\begin{eqnarray}
\rho_{T}&=&\frac{1}{2}(2T f_{T}-f-T), \label{4}\\
p_{T}&=&-\frac{1}{2}\left[-8\dot{H}T
f_{TT}+(2T-4\dot{H})f_{T}-f+4\dot{H}-T\right],\label{5}
\end{eqnarray}
and
\begin{equation}
T=-6H^{2} \label{6}
\end{equation}
For the choice of scale factor in Eq. (\ref{a}) we get from (\ref{2}) and (\ref{3}) that
\begin{equation}\label{rhoT}
\rho_T=\frac{1}{2}\left[t \dot{f}+\frac{6n^2}{t^2}-f\right]
\end{equation}
\begin{equation}\label{pT}
p_T=\frac{1}{6 n t^2}\left[-18 n^3-t^3 \dot{f} +12 n^2 (1+3 \dot{f}+t \ddot{f})+3 n t^2 (f+t \dot{f})\right]
\end{equation}
In the following subsections we shall derive solutions for reconstructed $f(T)$ in two ways. In Case I we shall get solution for $f(T)$ based on Eqs. (\ref{2}) and (\ref{rhoT}) and in Case II we shall reconstruct $f(T)$ based on Eqs. (\ref{3}) and (\ref{pT}).

\subsection{Case I}
In order to reconstruct $f(T)$ from the QCD GDE we consider in (\ref{2}) that $\rho=\rho_{gde}$ and we get
\begin{equation}\label{diffe}
3H^2-\rho_{gde}=\rho_T
\end{equation}
which is a linear differential equation with $t$ and $f$ as the independent and dependent variables respectively. Solving (\ref{diffe}) we can get $f(T)$ as a function of $t$
\begin{equation}
\begin{array}{c}\label{fre}
f(T)=\frac{1}{t^2}\left[4 n+C_1 t+(-1+2 n) t \alpha~log_{e}t\right]
\end{array}
\end{equation}
\begin{figure}
 \includegraphics[width=20pc]{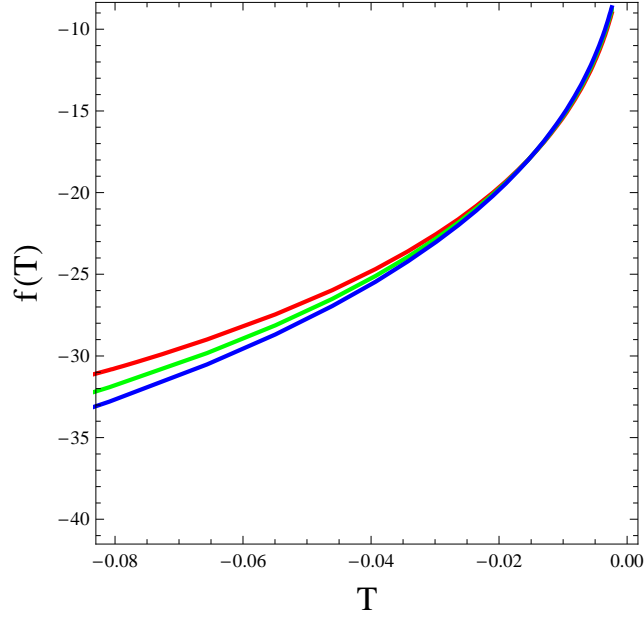}
\caption{\label{fig1} Reconstructed $f(T)\rightarrow 0$ as $T\rightarrow 0$ in Case I. This indicates a realistic model. }
\end{figure}
\begin{figure}
 \includegraphics[width=20pc]{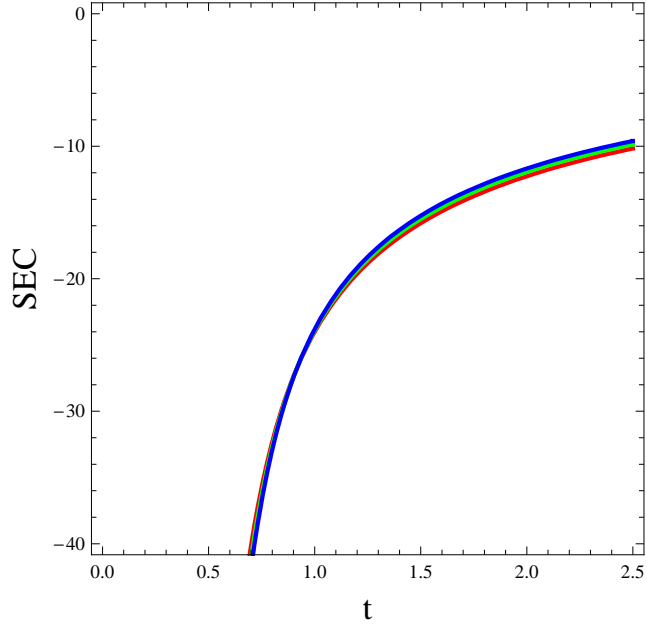}
\caption{\label{fig3} Strong energy condition $(\rho_T+\rho_{gde})+3(p_T+p_{gde})\geq 0$ is violated in Case I. Thus, $w_{eff}<-1/3$. which is consistent with the accelerated expansion of the universe.}
\end{figure}
\begin{figure}
 \includegraphics[width=20pc]{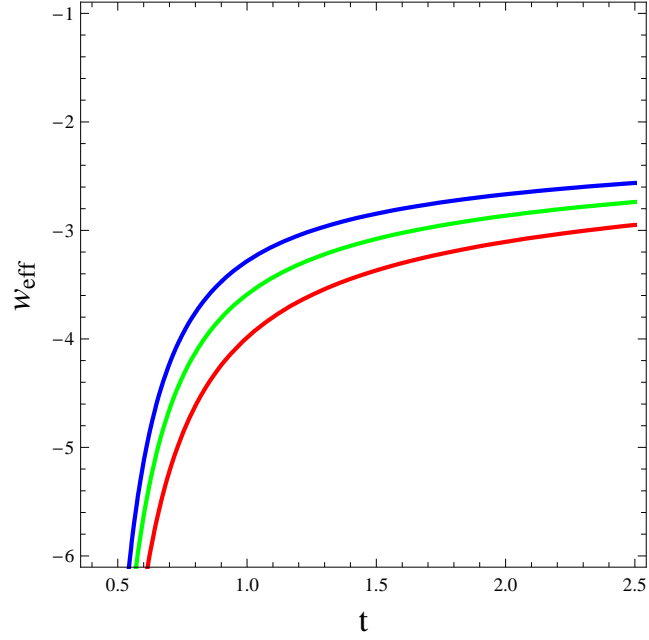}
\caption{\label{fig2} Reconstructed effective equation of state parameter $w_{eff}$ based on Eq. (\ref{wre}) in Case I. It is observed that $w_{eff}<-1$, which indicates phantom-like behaviour.}
\end{figure}

\begin{figure}
 \includegraphics[width=20pc]{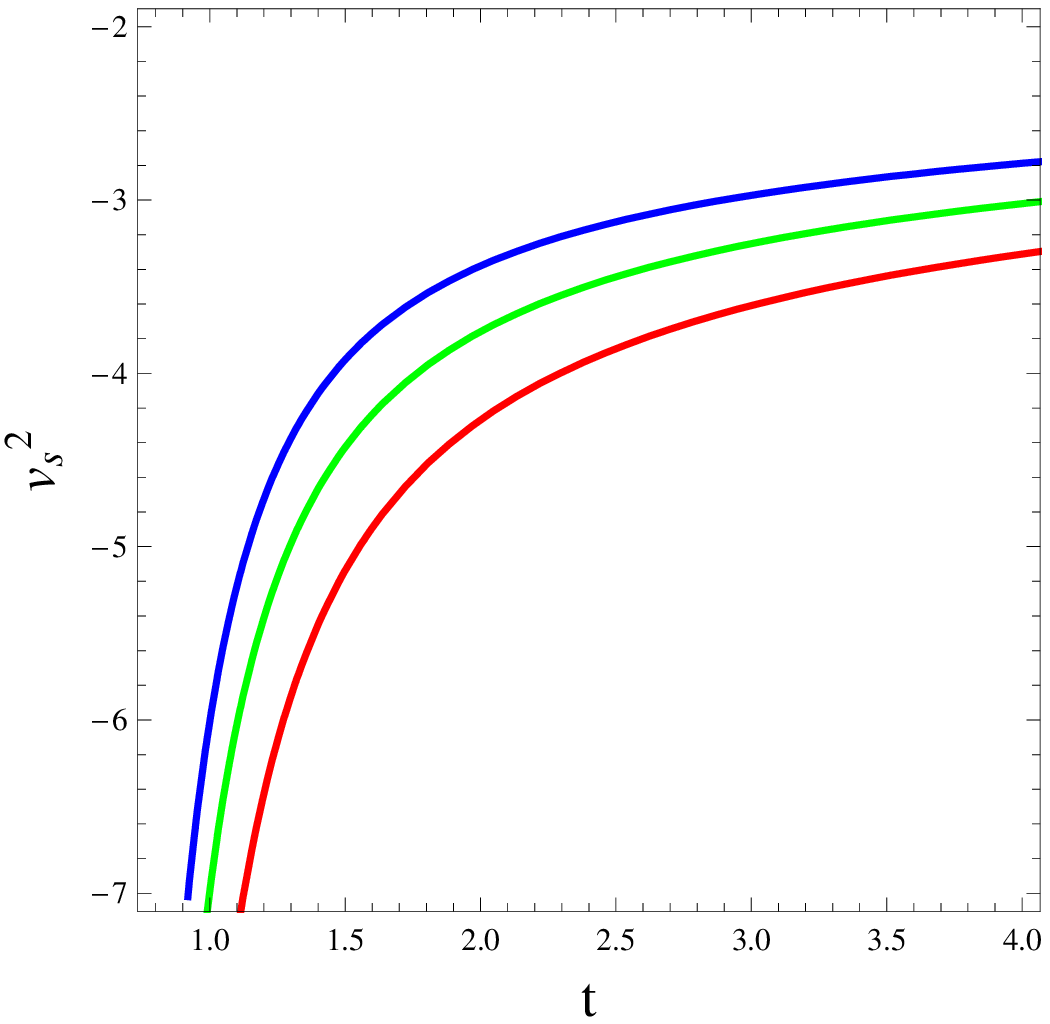}
\caption{\label{fig4} Squared speed of sound $v_s^2$ in Case I.}
\end{figure}
Subsequently using (\ref{rhoT}) and (\ref{pT}) we have reconstructed $\rho_T$  and $p_T$ as
\begin{equation}\label{rhore}
\begin{array}{c}
\rho_{T}=\left[-8 n t^3-C_1 t^4+6 n^2 t \left(2 C_1+t^2+2 \alpha \right)-\right.\\
\left.24 n^3 (-4+t \alpha )+t \left(-12 n^2+24 n^3+t^3-2 n t^3\right)
\alpha~log_{e}t\right]\\
\times(2 t^5)^{-1}
\end{array}
\end{equation}
\begin{equation}\label{pre}
\begin{array}{c}
p_{T}=\left[C_1 t^7+4 n t^4 \left(C_1+t^2+\alpha \right)-288 n^4 (-4+t \alpha )-\right.\\
\left.2 n^2 t^3 \left(-16+6 C_1 t+3 t^3+10 t \alpha
\right)+24 n^3 t \left(4 C_1-4 t^2+6 \alpha +t^3 \alpha \right)+\right.\\
\left.t \left(192 n^4+20 n^2 t^3-t^6+2 n t^3 \left(-2+t^3\right)-24 n^3 \left(4+t^3\right)\right)
\alpha~log_{e}t\right]\\
\times(2 t^8)^{-1}
\end{array}
\end{equation}
Finally, we get the reconstructed effective equation of state parameter $w_{eff}=\frac{p_T+p_{gde}}{\rho_T+\rho_{gde}}$ as
\begin{equation}
\begin{array}{c}\label{wre}
w_{eff}=\left[-t^7 \alpha +n t^7 (3 C_1+8 \alpha )-288 n^5 (-96+23 t \alpha )-12 n^3 t^3 \left(-40+12 C_1 t+3 t^3+25 t \alpha \right)+\right.\\
\left.12 n^2 t^4 \left(4 C_1+2 t^2+5 \alpha -t^3 \alpha \right)+72 n^4 t \left(28 C_1-20 t^2+46 \alpha +5 t^3 \alpha \right)+\right.\\
\left.3 n t \left(1344
n^4+80 n^2 t^3-t^6+2 n t^3 \left(-8+t^3\right)-96 n^3 \left(7+t^3\right)\right) \alpha  ~log_{e}t\right]\\
\times\left[3 n t^3 \left(-t^4 (C_1+2
\alpha )+12 n^2 t \left(4 C_1+t^2+5 \alpha \right)-120 n^3 (-4+t \alpha )+4 n t^3 (-4+t \alpha )+\right.\right.\\
\left.\left.t \left(-48 n^2+96 n^3+t^3-2 n t^3\right)
\alpha  ~log_{e}t\right)\right]^{-1}
\end{array}
\end{equation}
We now want to study an important quantity, namely the squared speed of sound, defined as:
\begin{equation}
v_{s}^{2}=\frac{\dot{p}}{\dot{\rho}}, \label{finaleqn}
\end{equation}
In our reconstruction problem, we shall have $v_{s}^{2}=\frac{\dot{p}_T}{\dot{\rho}_T+\dot{\rho}_{gde}}$.
The sign of $v_{s}^{2}$ is crucial for determining the stability
of a background evolution. Using Eqs. (\ref{qdeH}),(\ref{rhore}) and (\ref{pre}) the squared speed of sound has the following expression:
\begin{equation}
\begin{array}{c}
v_s^2=\left[t^7 (C_1+\alpha )-96 n^4 (-96+23 t \alpha )-4 n^2 t^3 \left(-40+12 C_1 t+3 t^3+25 t \alpha \right)+\right.\\
\left.2 n t^4 \left(8 C_1+4
t^2+10 \alpha -t^3 \alpha \right)+\right.\\
\left. 24 n^3 t \left(28 C_1-20 t^2+46 \alpha +5 t^3 \alpha \right)+t \left(1344 n^4+80 n^2 t^3-t^6+2 n t^3 \left(-8+t^3\right)-96
n^3 \left(7+t^3\right)\right) \alpha  ~log_{e}t\right]\\
\times\left[t^3 \left(-t^4 (C_1+2 \alpha )+12 n^2 t \left(4 C_1+t^2+5 \alpha \right)-120
n^3 (-4+t \alpha )+\right.\right.\\
\left.\left.4 n t^3 (-4+t \alpha )+t \left(-48 n^2+96 n^3+t^3-2 n t^3\right) \alpha  ~log_{e}t\right)\right]^{-1}
\end{array}
\end{equation}
Interpretations on the plots would be presented in the subsequent section.

\subsection{Case II}
As $p_{gde}=w_{gde} \rho_{gde}$ we have from conservation equation
 \begin{equation}\label{omega}
 3w_{gde}\Omega_{gde}=-\Omega_{gde}-\frac{\dot{\rho}_{\Lambda}}{9H^3}
 \end{equation}
 where $\Omega_{gde}=\frac{\rho_{gde}}{3H^2}$. Using Eq. (\ref{omega}) in (\ref{pT}) we have the following differential equation
 \begin{equation}\label{case2}
\frac{24 n}{t^2}f_{TT}+\left(6-\frac{2}{n}\right)f_T+\frac{t^2}{2 n^2}=\frac{6n^2-5n+1}{6n^3}\alpha t
 \end{equation}
 where,
 \begin{eqnarray}
 f_T=\frac{\dot{f}}{\dot{T}}=\frac{\dot{f}}{12 n^2}t^3\nonumber\\
 f_{TT}=\frac{\dot{f}_T}{\dot{T}}=\frac{t^5(3\dot{f}+t\ddot{f})}{144 n^4}\nonumber
 \end{eqnarray}
 The over-dot indicates derivative with respect to $t$. Solving (\ref{case2}) we get the reconstructed $f(T)$ as
\begin{equation}
\begin{array}{c}\label{fre2}
f(T)=\frac{1}{t^2\sqrt{\pi }}2^{-\frac{5}{2}-\frac{3 n}{2}} 3^{-\frac{1}{4}-\frac{3 n}{4}}\\
 \left(\frac{\sqrt{n^3}}{t}\right)^{-\frac{1}{2}-\frac{3
n}{2}} \left(-3^{\frac{1}{2}+
\frac{3 n}{2}} \frac{6n^2-5n+1}{6n^3}\alpha~\pi  \left(\frac{\sqrt{n^3}}{t}\right)^{1+3 n} t~pFq\left[\left\{\frac{1}{2}\right\},\left\{\frac{3
(1+n)}{2},\frac{3}{2}\right\},\frac{3 n^3}{4 t^2}\right]\right.\\
\left.\left(2 BesselK\left[\frac{1}{2}+\frac{3 n}{2},\frac{\sqrt{3} \sqrt{n^3}}{t}\right]+\pi
 BesselI\left[\frac{1}{2}+\frac{3 n}{2},\frac{\sqrt{3} \sqrt{n^3}}{t}\right] \text{Sec}\left[\frac{3 n \pi }{2}\right]\right)+2 t \left(2
\left(\frac{1}{t}\right)^{3 n}\right.\right.\\
\left.\left. \left(BesselK\left[\frac{1}{2}+\frac{3 n}{2},\frac{\sqrt{3} \sqrt{n^3}}{t}\right] C_1+2^{1+3 n} \sqrt{\pi
} BesselI\left[\frac{1}{2}+\frac{3 n}{2},\frac{\sqrt{3} \sqrt{n^3}}{t}\right] C_2 Gamma\left[\frac{3 (1+n)}{2}\right]\right)+\right.\right.\\
\left.\left.8^n \frac{6n^2-5n+1}{6n^3}\alpha~
\pi ^{3/2} BesselI\left[\frac{1}{2}+\frac{3 n}{2},\frac{\sqrt{3} \sqrt{n^3}}{t}\right] Gamma\left[-\frac{3 n}{2}\right]\right.\right.\\ \left.\left.pFq\left[\left\{-\frac{3
n}{2}\right\},\left\{\frac{1}{2}-\frac{3 n}{2},1-\frac{3 n}{2}\right\},\frac{3 n^3}{4 t^2}\right] Sec\left[\frac{3 n \pi }{2}\right]\right)\right)
\end{array}
\end{equation}
Because of the huge and complicated forms of the algebraic expressions, we shall create plots of SEC, $w_{eff}$ and $v_s^2$ instead of showing their expressions. The plots are interpreted in the subsequent section.
\begin{figure}
 \includegraphics[width=20pc]{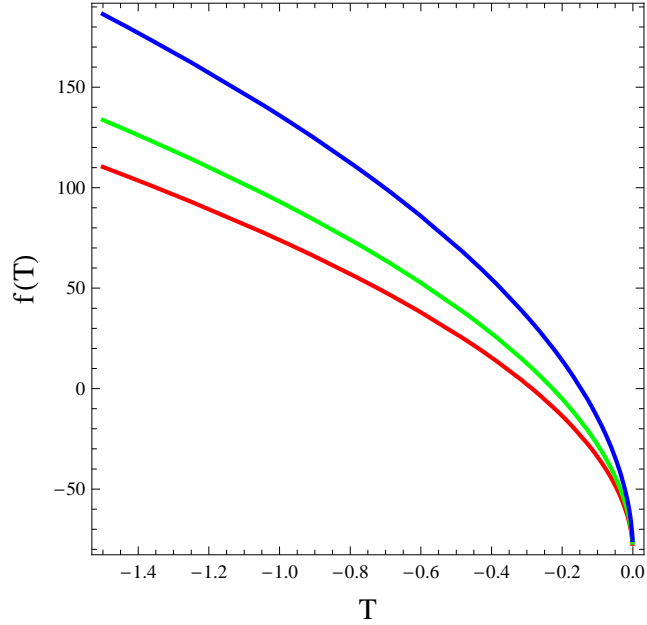}
\caption{\label{fig5} Reconstructed $f(T)\rightarrow 0$ as $T\rightarrow 0$ in Case II. }
\end{figure}

\begin{figure}
 \includegraphics[width=20pc]{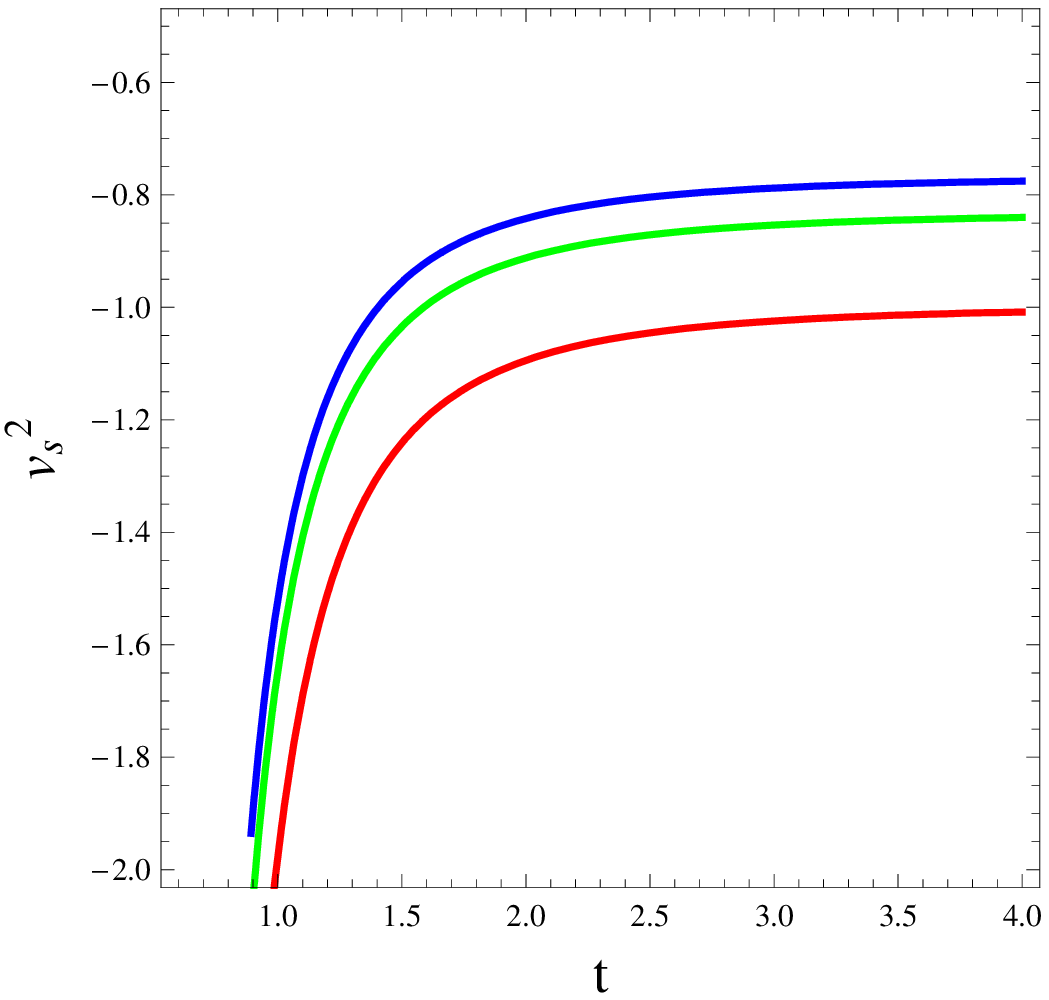}
\caption{\label{fig6} Squared speed of sound $v_s^2$ in Case II. }
\end{figure}

\begin{figure}
 \includegraphics[width=20pc]{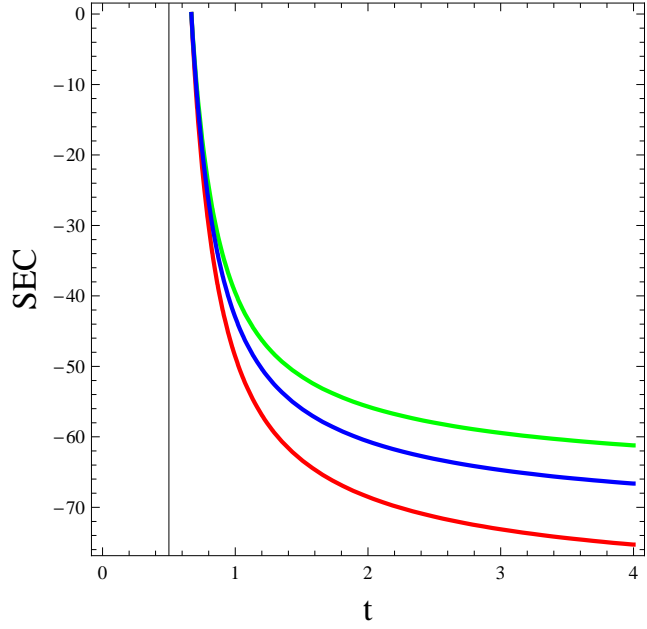}
\caption{\label{fig7} The SEC in Case II. }
\end{figure}

\begin{figure}
 \includegraphics[width=20pc]{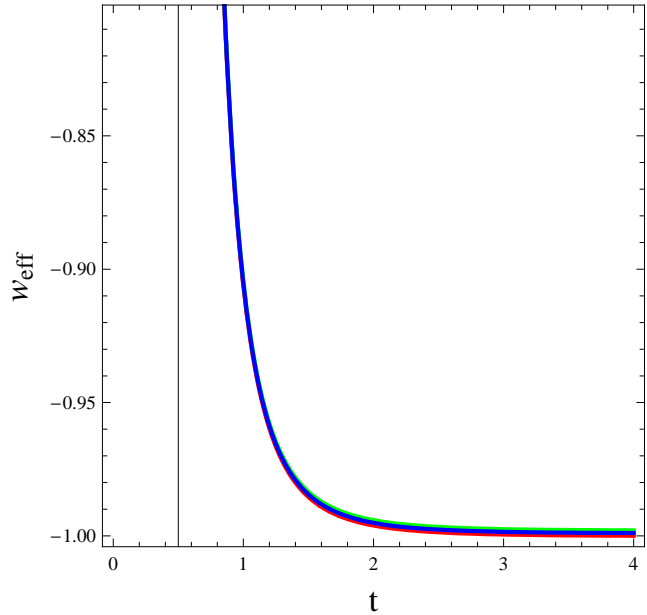}
\caption{\label{fig8} The effective equation of state parameter $w_{eff}$ in Case II. }
\end{figure}

\subsection{Discussion on the plots}
In this subsection we discuss the consequences of the above reconstruction scheme through plots.
\begin{description}
  \item[Case I:] In this case, the reconstructed $f(T)$ is given in the Eq. (\ref{fre}). Based on this equation, we have reconstructed $p_T$ and $\rho_T$ and we have used them to study the cosmological consequences of the reconstructed $f(T)$. In the plots corresponding to Case I we have taken $C_1=-18,~n=1/20$. In all of the plots, red, green and blue lines corresponds to $\alpha=5.6,~5$ and $4.5$ respectively. In Fig. 1 we have plotted the reconstructed $f(T)$ as derived in Eq. (\ref{fre}) against torsion $T$ as given in (\ref{6}). It is apparent from the figure that $f(T)\rightarrow 0$ as $T\rightarrow 0$. It has been discussed in reference \cite{recons3} that satisfaction of the above condition is a sufficient condition for a realistic model. In Fig. 2 we examine validity of strong energy condition (SEC) with the evolution of the universe and it is observed that in the late stage of the universe $\rho_T+\rho_{gde}+3(p_T+p_{gde})< 0$ (based on Eqs. (\ref{rhore}) and (\ref{pre})) and it indicates violation of SEC in the late stage of the universe. This implies that $w_{eff}\leq -1/3$ and hence it is consistent with DE property and accelerated expansion of the universe.  In Fig. 3, we plot the effective equation of state parameter $w_{eff}=\frac{p_T+p_{gde}}{\rho_T+\rho_{gde}}$ based on Eq. (\ref{wre}). It is always observed that $(w_{eff}<-1)$ phase and there is no apparent possibility of reaching the phantom boundary of $-1$. Thus, the effective equation of state parameter behaves like phantom. Finally, in figure 4 we have plotted the squared speed of sound $v_s^2$ and it is found that $v_s^2<0$ and this indicates that the model is classically unstable.\\
  \item[Case II:] In this case, the reconstructed $f(T)$ is given in the Eq. (\ref{fre2}). We have reconstructed $p_T$ and $\rho_T$ to see the cosmological consequences of (\ref{fre2}) that is plotted against $T$ in Fig. 5. In all the figures under Case II we have taken $n=1/20,~C_1=-2$ and $C_2=2$. Red, green and blue lines corresponds to $\alpha=3.3,~3.5$ and $3.8$ respectively. It is observed in Fig. 5 that $f(T)\rightarrow 0$ as $T\rightarrow 0$, although the pattern is different from Case I. Thus, like (\ref{fre}), a realistic model is represented also by (\ref{fre2}). The squared speed of sound presented in Fig. 6 exhibits the same property as Case I i.e. $v_s^2<0$, thus indicating a classical instability of the model. Strong energy condition as presented in Fig. 7 is violated by the model like Case I and hence $w_eff<-1/3$ that is consistent with the accelerated expansion of the universe. The models in Case I and Case II get their difference in the effective equation of state parameter $w_{eff}$ plotted in Fig. 8, where it is observed that initially $w_{eff}>-1$ and with evolution of the universe it is approaching towards $w_{eff}=-1$ i.e. the phantom boundary. Although $w_{eff}$ is behaving like ``quintessence", it has a tendency of approaching towards phantom phase of the universe.
\end{description}

\section{Concluding remarks}
In the present work we have studied a reconstruction scheme for $f(T)$ gravity based on QCD ghost dark energy. In the modified field equations we have considered $\rho$ as the $\rho_{gde}$ in a flat universe with power law form of the scale factor. Because of the choice of the scale factor, the $\rho_{gde}$ could be expressed as a function of $t$. Subsequently, considering the two field equations, we have reconstructed $f(T)$ in two forms described as Case I and Case II respectively and given in Eqs. (\ref{fre}) and (\ref{fre2}) respectively. In both of the cases $f(T)\rightarrow 0$ as $T\rightarrow 0$, that indicates a realistic model in both of the cases. Since both of the forms appear as functions of $t$, we could get their time-derivatives and could successfully reconstruct the density $\rho_T$ and pressure $p_T$ contributions due to torsion $T$. Using these reconstructed $\rho_T$ and $p_T$ we could generate effective equation of state parameter $w_{eff}$ in both of the cases and we observed that in Case I, $w_{eff}<<-1$ and in Case II, $w_{eff}\geq -1$. Thus Case I and Case II generates ``phantom" and ``quintessence"-like $w_{eff}$ respectively. One prominent different was that for Case I we are staying far below the phantom boundary and in Case II it is getting asymptotic at the phantom boundary coming from $w_{eff}>-1$. Based on this difference of the behaviours of $w_{eff}$ it may be interpreted that Case II i.e. Eq. (\ref{fre2}) represents a more acceptable model at is can show an approach towards phantom phase of the universe starting from quintessence. Due to non-positivity of the squared sound speed $v_s^2$ as seen in the plots, both of the QCD ghost $f(T)$ models are classically unstable against perturbations in flat and non-flat Friedmann-Robertson-Walker backgrounds. This instability problem is consistent with result presented for QCD ghost dark energy model by \cite{QCD1}. However, instability problem raised by negativity of $v_s^2$ by arguing that the Veneziano ghost does not have a physical propagating degree of freedom and the corresponding GDE model does not violate unitarity causality or gauge invariance. This argument can be seen in \cite{plb}.

We would like to mention the work of \cite{nojiri45}, where the dark energy universe equation of state with inhomogeneous, Hubble parameter dependent
term was considered and crossing of the phantom barrier was realized. In our current work we have reconstructed $f(T)$ gravity based on QCD ghost dark energy and our equation of state parameter has been found to be above $-1$ and gradually tending to $-1$. We propose as future work to consider the assumed equation of state parameter of the work of \cite{nojiri45} in $f(T)$ reconstruction and to investigate whether this helps the reconstructed $f(T)$ to cross the phantom barrier.

\section{Acknowledgements}
Sincere thanks are due to the anonymous reviewer for constructive suggestion. Financial support from the Department of Science and Technology (Govt. of India) under Project Grant No. SR/FTP/PS-167/2011 is duly acknowledged.

\end{document}